\documentclass[final,5p,times,twocolumn,letterpaper]{elsarticle}

\usepackage{natbib}
\usepackage{cancel}
\usepackage{ulem}
\usepackage{amsthm}         
\usepackage{amsmath} 	   
\usepackage{amssymb}       
\usepackage{amsfonts}
\usepackage{graphicx} 	    
\usepackage{verbatim}
\usepackage{color}
\usepackage{float}

\definecolor{dark-red}{rgb}{0.0,0.0,0.0}
\definecolor{dark-green}{rgb}{0.0,0.0,0.0}
\definecolor{dark-blue}{rgb}{0.0,0.0,0.0}

\begin{document}

\title{Inflationary paradigm in trouble after {Planck}2013}

\author[add1,add2]{Anna Ijjas}
\ead{aijjas@cfa.harvard.edu}
\author[add1,add3,add4]{Paul J. Steinhardt\corref{cor1}}
\ead{steinh@princeton.edu}
\author[add1]{Abraham Loeb}
\ead{aloeb@cfa.harvard.edu}

\address[add1]{Harvard-Smithsonian Center for Astrophysics, Cambridge, MA 
02138, USA}
\address[add2]{University Observatory Munich, 81679 Munich, Germany}
\address[add3]{Department of Physics, Princeton University, Princeton, NJ 
08544, 
USA}
\address[add4]{Princeton Center for Theoretical Science, Princeton 
University, Princeton, NJ 08544 USA}

\cortext[cor1]{Corresponding author}

\date{\today}

\begin{abstract} Recent results from the \textit{Planck} satellite combined with earlier observations from WMAP, ACT, SPT and other experiments eliminate a wide spectrum of more complex inflationary models and favor models with a single scalar field, as reported by the \textit{Planck} Collaboration.  More important, though, is that all the simplest inflaton models are disfavored statistically relative to those with plateau-like potentials.
We discuss how a restriction to plateau-like models has three independent serious drawbacks: it exacerbates both the initial conditions problem and the multiverse-unpredictability problem and it creates a  new difficulty that we call the inflationary ``unlikeliness problem.''
Finally, we comment on problems reconciling inflation with a standard model Higgs, as suggested by recent LHC results.  In sum, we find that recent experimental data disfavors all the best-motivated inflationary scenarios and introduces new, serious difficulties that cut to the core of the inflationary paradigm.  Forthcoming searches for B-modes, non-Gaussianity and new particles should be decisive.       
\end{abstract}

\maketitle

The Planck satellite data reported in 2013 \cite{Ade:2013rta}  shows with high precision that we live in a remarkably simple universe.  The measured spatial curvature is small; the spectrum of fluctuations is nearly scale-invariant; there is a small spectral tilt, consistent with there having been a simple dynamical mechanism that caused the smoothing and flattening; and the fluctuations are nearly Gaussian, eliminating exotic and complicated dynamical possibilities, such as inflationary models with non-canonical kinetic energy and multiple fields. (In this Letter, we will not discuss the marginal deviations from isotropy on large scales reported by the \textit{Planck} Collaboration \cite{Ade:2013nlj}.)  The results not only impose tight quantitative constraints on all cosmological parameters \cite{Ade:2013lta}, but, qualitatively, they call for a cosmological paradigm whose simplicity and parsimony matches the nature of the observed universe.

The \textit{Planck} Collaboration attempted to make this point by describing the data as supporting the \textit{simplest} inflationary models \cite{Guth:1980zm,Linde:1981mu,Albrecht:1982wi}.  However, the models most favored by their data (combined with earlier results from WMAP, ACT, SPT and other observations \cite{Sievers:2013wk}) are simple by only one criterion: an inflaton potential with a single scalar field suffices to fit the data. By several other important criteria described in this Letter, the favored models are \textit{anything but simple}:  Namely, they suffer from exacerbated forms of initial conditions and multiverse problems, and they create a new difficulty that we call the inflationary ``unlikeliness problem.''  That is, the favored inflaton potentials are exponentially unlikely according to the logic of the inflationary paradigm itself. The unlikeliness problem arises even if we assume ideal initial conditions for beginning inflation, ignore the lack of predictive power stemming from eternal inflation and the multiverse, and make no comparison with alternatives.  Thus, the three problems are all independent, all emerge as a result of the data, and all point to the inflationary paradigm encountering troubles that it did not have before. We further speculate about how recent results from the Large Hadron Collider (LHC) suggesting a standard model Higgs could create yet another problem for inflation.

Our analysis is based on considering the ``favored'''' models according to the current observations. (Here and throughout the Letter we use the ranking terminology of the \textit{Planck} Collaboration).  Although the simplest inflationary models are ``disfavored'''' relative to these by 1.5~$\sigma$ or more, it is too early in some cases to declare them ``ruled out.''"   We discuss in the conclusions how forthcoming searches for B-modes, non-Gaussianity and new particles could amplify, confirm, or resolve the problems for inflation.

\section{Which inflationary models survive after Planck2013?}
Planck2013 has added impressively to previous results in three ways.  First, it has shown that the non-Gaussianity is small.  This eliminates a wide spectrum of more complex inflationary models and favors models with a single scalar field.  This restriction to single-field models is what justifies focusing on the plot of $r$ (the ratio of tensor to scalar fluctuations) versus $n_s$ (the scalar spectral index), since it is optimally designed to discriminate among the single-field possibilities.  
In terms of the $r$-$n_s$ plot, a second contribution of Planck2013 \cite{Ade:2013rta} has been to independently confirm the results obtained previously by combining WMAP with other observations.   The data disfavors by 1.5$\sigma$ or more all the simplest inflation models: power-law potential and chaotic inflation \cite{Linde:1983gd}, exponential potential and power-law inflation \cite{Lucchin:1984yf}, inverse power-law potential \cite{Barrow:1990vx,Muslimov:1990be}.  Third, the $r$-$n_s$ plot favors instead a special subclass of inflationary models with \textit{plateau-like} inflaton potentials. These models -- simple symmetry breaking \cite{Linde:1981mu,Albrecht:1982wi,Olive:1989nu},  natural (axionic) \cite{Freese:1990rb}, symmetry breaking with non-minimal (quadratic) coupling \cite{Salopek:1988qh,Fakir:1990eg,Bezrukov:2007ep}, $R^2$ \cite{Starobinsky:1980te}, hilltop \cite{Steinhardt:1984jj} -- are simple in the sense that they all can be formulated (in some cases via changes of variable \cite{Maeda:1988ab,Wands:1993uu,Faulkner:2006ub,Rodrigues:2011zi}) as  single-field, slow-roll models with a canonical kinetic term in the framework of Einstein gravity \cite{Ade:2013rta}.  A distinctive feature of this subclass of models, following from the Planck2013 constraint on $r$ ($r_{0.002} < 0.12$ at 95\% CL), that will be important in our analysis is that the energy scale of the plateau ($M_{\text{I}}^4$) is at least 12 orders of magnitude below the Planck scale $\sim M_{\text{Pl}}^4$ \cite{Ade:2013rta},
\begin{equation}
M_{\text{I}}^4 \lesssim  \frac{3\pi^2A_s}{2}\,r\,M_{\text{Pl}}^4\sim 10^{-12}\,M_{\text{Pl}}^4\,\frac{r_{\ast}}{0.12}
\end{equation}
at 95\% CL, where $A_s$ is the scalar amplitude and $r_{\ast}$ the value of  $r$ evaluated at Hubble exit during inflation of mode with wave number $k_{\ast}$.

A classic example that we will consider first is the original \textit{new inflation} model \cite{Linde:1981mu,Albrecht:1982wi} based on  a Higgs-like inflaton, $\phi$, and potential $V(\phi)= \lambda (\phi^2-\phi_0^2)^2$, as illustrated in Fig.\,1a.  The plateau region is the range of small $\phi \ll \phi_0$.  Other examples illustrated in Figs.\,1b and 1c will then be considered. 

\begin{figure}
\begin{center}
\includegraphics[scale=0.575]{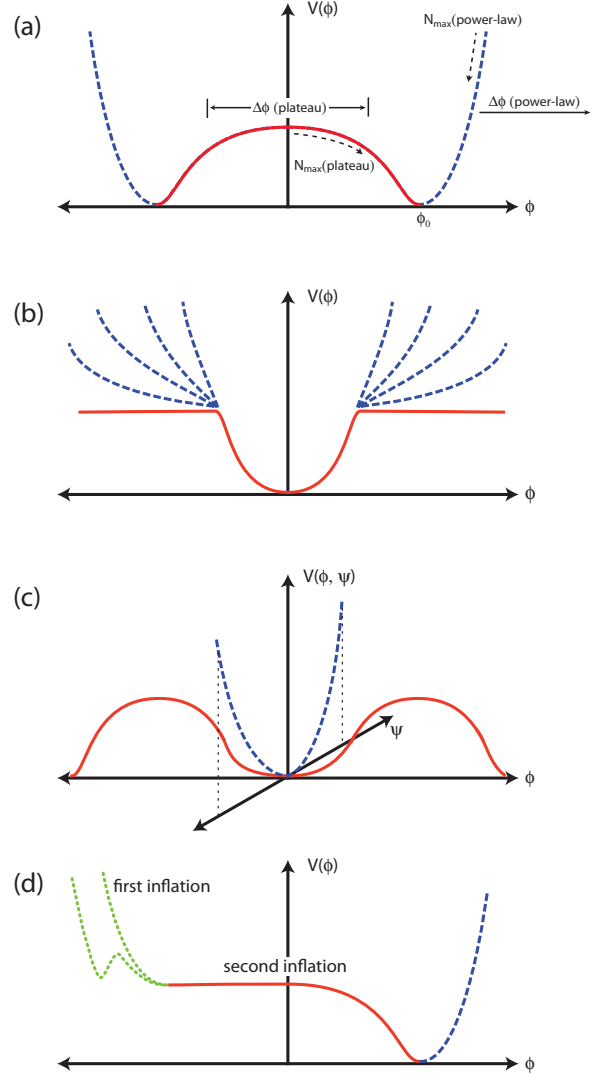}
\caption{Plateau-like models favored by Planck2013 data: (a) Higgs-like potential $V$ with standard Einstein gravity that has both plateau at $\phi \ll \phi_0$ (solid red) and power-law behavior at $\phi \gg \phi_0$ (dashed blue), where $N_{\text{max}}$ is the maximum number of e-folds of inflation possible for the maximal range $\Delta \phi$;  (b) unique plateau-like model (solid red) for semi-infinite range of $\phi$ if perfectly tuned compared to continuum of power-law inflation models (dashed blue) without tuning; (c) periodic (axion-like) plateau potential (solid red) for $\phi$ plus typical power-law inflation potential (dashed blue) for second field $\psi$; (d) designed inflationary potential with power-law inflation segment or false vacuum segment (dotted green) grafted onto a  plateau model (solid red).}
\end{center}
\end{figure}

An obvious difference between plateau-like models like this and the simplest inflationary models, like $V(\phi)= \lambda \phi^4$, is that the simplest models require only one parameter and  absolutely no tuning of parameters to obtain 60 or more e-folds of inflation while the plateau-like models require three or more parameters and must be fine-tuned to obtain even a minimal amount of inflation.  For $V(\phi)= \lambda \phi^4$ all that is required is that $\phi \ge M_{\text{Pl}}$, where $M_{\text{Pl}}$ is the Planck mass.  However, the fine-tuning of parameters is a minor issue within the context of the more serious problems described below that undercut the inflationary paradigm altogether. 

\section{How do plateau-like inflationary models affect the initial conditions problem?}
As originally imagined, inflation was supposed to smooth and flatten the universe beginning from arbitrary initial conditions after the big bang \cite{Guth:1980zm}.  However, this view had to be abandoned as it was realized that large inflaton kinetic energy and gradients within a Hubble-sized patch prevent inflation from starting.  While some used statistical mechanical reasoning to argue that the initial conditions required for inflation are exponentially rare \cite{Penrose:1988mg,Gibbons:2006pa}, the almost ``universally accepted'' \cite{Liddle:2000cg} assumption for decades, originally due to Linde \cite{Linde:1983gd,Linde:1984ir,Linde:1984st,Linde:1985ub,Kofman:1985aw,Belinsky:1987gy,Toporensky:1999pk,Pavluchenko:2001mt,Kofman:2002cj,Linde:2005ht}, has been that the natural initial condition when the universe first emerged from the big bang and reached the Planck density is having all different energy forms of the same order.  For the inflaton, this means $\frac{1}{2}\dot{\phi}^2 \sim \frac{1}{2}(\partial_i\phi)^2 \sim V(\phi) \sim M_{\text{Pl}}^4$. Roughly speaking, the assumption is based on the notion that all these forms of energy density span the same range, from zero to $M_{\text{Pl}}^4$, so it is plausible to have them of the same order at a time when the total energy density is $M_{\text{Pl}}^4$. Evolving forward in time from these initial conditions, $V(\phi)$ almost immediately comes to dominate the energy density and triggers inflation before the kinetic and gradient energy can block it from starting. 

 
After Planck2013, the very same argument used to defend inflation now becomes a strong argument against it. Because the potential energy density of the plateau $M_{\text{I}}^4$ is bounded above and must be at least a trillion times smaller than the Planck density to obtain the observed density fluctuation amplitude, the only patches that exist have $\frac{1}{2}\dot{\phi}^2 \sim \frac{1}{2}(\partial_i\phi)^2 \gg V(\phi)$.  In particular, beginning from these revised initial conditions and evolving forward in time, the kinetic energy decreases as $1/a^6$ and the gradient energy as $1/a^2$, where $a(t)$ is the Friedman-Robertson-Walker scale factor.  Hence, beginning from roughly equal kinetic and gradient energy, gradients and inhomogeneities quickly dominate and the combination blocks inflation from occurring.  

To quantify the problem, for inflation to initiate, there must be a seed region at the Planck density ($t=t_{\text{Pl}}$) that remains roughly homogeneous until inflation begins ($t=t_{\text{I}}$) and whose radius $r(t)$ has expanded to a size  at least equal to a Hubble radius, $H^{-1}(t_{\text{I}})$ at the time inflation initiates.  After Planck2013, this requires, by simple comparison of the scales $M_{\text{Pl}}/M_{\text{I}} \sim 10^3\cdot(10^{16}\,\text{GeV}/M_{\text{I}})$ as constrained by Planck2013, that there exist homogeneous initial volumes before inflation begins whose size is  
\begin{eqnarray}
r^3(t_{\text{Pl}}) &\gtrsim& \left[ a(t_{\text{Pl}}) \int^{t_{\text{I}}}_{t_{\text{Pl}}} \frac{d\,t}{a}\right]^{3} \sim \left[\frac{a(t_{\text{Pl}})\,H(t_{\text{Pl}})}{a(t_{\text{I}})\,H(t_{\text{I}})}H^{-1}(t_{\text{Pl}})\right]^{3}\nonumber \\ 
&>& 10^{9} \left( \frac{10^{16}\,\text{GeV}}{M_{\text{I}}} \right)^3  H^{-3}(t_{\text{Pl}}) \,,
\end{eqnarray}
-- initial smoothness on the scale of a billion or more Hubble volumes \cite{Liddle:2000cg}!
 
In sum, by favoring only plateau-like models, the Planck2013 data creates a serious new challenge for the inflationary paradigm: the universally accepted assumption about initial conditions no longer leads to inflation; instead, inflation can only begin to smooth the universe if the universe is unexpectedly smooth to begin with!

\section{Is a plateau-like potential likely according to the inflationary paradigm?}
All inflationary potentials are not created equal. The odd situation after Planck2013 is that inflation is only favored for a special class of models that is exponentially unlikely according to the inner logic of the inflationary paradigm itself.  The situation is independent of the initial conditions problem described above;  even assuming ideal conditions for initiating inflation, the fact that only plateau-like models are favored is paradoxical because inflation requires more tuning, occurs for a narrower range of parameters, and produces exponentially less plateau-like inflation than the now-disfavored models with power-law potentials.  This is what we refer to as the inflationary ``unlikeliness problem.''

To illustrate the problem, we continue with the classic plateau-like model $V(\phi) = \lambda(\phi^2-\phi_0^2)^2$. Like most plateau-like inflationary models, the plateau terminates at a local minimum, and then the potential grows as a power-law ($\sim \lambda \phi^4$ in this case) for large $\phi$.  The problem arises because \textit{within this scenario} the same minimum can be reached in two different ways, either by slow-roll inflation along the plateau or by slow-roll inflation from the power-law side of the minimum.  
It is easy to see that inflation from the power-law side requires less tuning of parameters, occurs for a much wider range of $\phi$, and produces exponentially more inflation:
constraints on an inflationary model are determined by the amount of inflation ($N \sim 60$); the scale of density fluctuations ($\delta \rho / \rho \sim 10^{-5}$); and the condition called ``graceful exit'' (which ensures that inflation ends locally and marks the start of reheating).  Using the well-known slow-roll approximation,  $N \sim V / V''$, $d \rho / \rho \sim V^{3/2}/V'$, these constraints can be specified for both plateau-like $\sim \lambda\phi_0^4-2\lambda\phi_0^2\phi^2$ and power-law $\sim \lambda \phi^4$ inflation \cite{Mukhanov:2005sc}. 

One immediately observes that the first constraint imposes no parameter tuning constraints on power-law models but does require fine-tuning for plateau-like models.  For the plateau-like model, inflation occurs if $\phi$ lies in the range \begin{equation}
\Delta \phi(\text{plateau}) \lesssim \phi_0 \sim M_{\text{Pl}},
\end{equation}  
and the maximum number of e-folds is
\begin{eqnarray}
N_{\text{max}} (\text{plateau}) &=& \int^{t_e}_{t_i} H\, d\,t \sim \frac{8\pi}{M_{\text{Pl}}^2} \int^{\phi_e}_{\phi_i} \frac{V}{V'}\, d\phi \nonumber\\ & \sim  & 8 \pi \phi_0^2/M_{\text{Pl}}^2 \,.
\end{eqnarray}
By comparison, coming from the power-law side of the same potential, inflation occurs for the range $\Delta \phi(\text{power-law}) \lesssim \lambda^{-1/4} M_{\text{Pl}}$, so that 
\begin{equation} 
\Delta \phi(\text{power-law}) \gg \Delta \phi(\text{plateau}),
\end{equation}
where we have followed convention in confining the power-law range to those values for $\phi$ for which $V(\phi)$ is less than the Planck density and used the fact that $\lambda$ must be of order $10^{-15}$ to obtain the observed density perturbation amplitude on large scales.   
Also, the maximum integrated amount of inflation on the power-law side is 
\begin{eqnarray}
N_{\text{max}}(\text{power-law}) & \sim & \text{max}\{8\pi (\phi^2_{\text{initial}} - \phi^2_{\text{end}}) / M_{\text{Pl}}^2\} \nonumber \\ & \sim & \lambda^{-1/2} N_{\text{max}}(\text{plateau})\nonumber\\ &\gg& N_{\text{max}}(\text{plateau}). 
\end{eqnarray}
Obviously, given the much larger field-range for $\phi$ and larger amount of expansion, inflation from the power-law side is exponentially more likely according to the inflationary paradigm; yet {Planck}2013 forbids the power-law inflation and only allows the unlikely plateau-like inflation.  This is what we call the inflationary unlikeliness problem.  

Although we have demonstrated the principle so far for only a single potential, completion of most scalar field potentials, plateau-like or not, entails power-law or exponential behavior at large values of $\phi$.  There are notable examples that have no power-law completion, such as axion and moduli potentials.  However, as discussed in Sec.~5, unless all scalar fields defining our vacuum are of this nature, inflation from a scalar field with power-law or exponential behavior is exponentially more likely; but this is disfavored by Planck2013. 

Therefore, post-Planck2013 inflationary cosmology faces an odd dilemma.  The usual test for a theory is whether experiment agrees with model predictions.  Obviously, inflationary plateau-like models pass this test.  However, this cannot be described as a success for the inflationary paradigm, since, according to inflationary reasoning, this particular class of models is highly unlikely to describe reality. The unlikeliness problem is an alarm warning us that a paradigm can fail even though observations favor a class of models if the paradigm predicts the class of models is unlikely.

\section{Is Planck2013 data compatible with the multiverse?}
A well-known property of almost all inflationary models is that, once inflation begins, it continues eternally producing a multiverse \cite{Steinhardt:1982kg,Vilenkin:1983xq} in which ``anything that can happen will happen, and it will happen an infinite number of times'' \cite{Guth:2007ng}.  A result is that all cosmological possibilities (flat or curved, scale-invariant or not, Gaussian or not,  \textit{etc.}) and any combination thereof are equally possible,  potentially rendering inflationary theory totally unpredictive.  Attempts to introduce a measure principle \cite{Garriga:2001ri,Garriga:2005av,Aguirre:2006ak,Vilenkin:2006xv,Winitzki:2006rn,Freivogel:2011eg} or anthropic principle \cite{Weinberg:1987dv,Weinberg:2005fh,Susskind:2003kw} to restore predictive power have met with difficulty.  For example, the most natural kind of measure, weighting by volume, does not predict our universe to be likely.  Younger patches \cite{Linde:1994gy,Guth:2000ka} and Boltzmann brains/babies \cite{Albrecht:2004ke,DeSimone:2008if} are exponentially favored.  

Planck2013 results lead to a new twist on the multiverse problem that is independent of the initial conditions and unlikeliness problems described above.  
The plateau-like potentials selected by Planck2013 are in the class of eternally inflating models, so the multiverse and its effects on predictions must be considered. In a multiverse, each measured cosmological parameter represents an independent test of the multiverse in the sense one could expect large deviations from any one of the naive predictions.  The more observables one tests, the greater the chance of many-$\sigma$ deviations from the naive predictions. Hence, it is surprising that the Planck2013 data agrees so precisely with the naive predictions derived by totally ignoring the multiverse and assuming purely uniform slow-roll down the potential.

\section{Is there any escape from these new problems?}
In the previous sections we introduced three independent problems stemming from the Planck2013 observations: a new initial conditions problem, a worsening multiverse-unpredictability problem, and a novel kind of discrepancy between data and paradigm that we termed the unlikeliness problem.  It is reasonable to ask: is there any easy way to escape these problems?

One approach that cannot work is the anthropic principle since the new problems discussed in this Letter all derive from the fact that Planck2013 disfavors the simplest inflationary potentials while there is nothing anthropically disadvantageous about those models or their predictions.

The multiverse-unpredictability problem has been known for three decades before Planck2013  and, thus far, lacks a solution.  For example, weighting by volume and bubble counting, the most natural measures by the inner logic of the inflationary paradigm, fail.    

By contrast, one might imagine the unlikeliness problem first brought on by Planck2013 could be evaded by a different choice of potential.  Above we used as an example the potential $V(\phi)= \lambda (\phi^2 - \phi_0^2)^2$, which has a plateau for $\phi \ll \phi_0$ and a power-law form for $\phi \gg \phi_0$.  Here it was clear that inflation from the power-law side is exponentially more likely because inflation occurs for a wider range of $\phi$ and generates exponentially more accelerated expansion.  

An alternative, in principle, is to have a plateau at large $\phi$ and no power-law behavior, as sketched in Fig.\,1b.  The problem with this is that the desired flat behavior, marked in red, is a unique form that only occurs for a precise cancellation order by order in $\phi$ (if one imagines $V$ expanded in a power series in $\phi$). 
Within the inflationary paradigm, this perfect cancellation is not only ultra-fine tuned, but also uncalled for since there are infinitely many power-law inflationary completions of the potentials (blue-dashed) in which $V$ increases as a power of $\phi$. The single plateau possibility is extremely unlikely compared to the continuum of blue-dashed possibilities. Yet now Planck2013 disfavors everything except for the unlikely plateau case.
Examples of this type include the Higgs inflationary model with non-minimal coupling $f(\phi)R$ with $f(\phi)= M_{Pl}^2 + \xi \phi^2$ \cite{Bezrukov:2007ep,Bezrukov:2009db,Bezrukov:2010jz,Linde:2011nh,GarciaBellido:2011de,Bezrukov:2012hx}  and the $f(R)= R + \xi R^2$ inflation model \cite{Starobinsky:1980te}, where $R$ is the Ricci scalar, once they are converted by changes of variable to a theory of a scalar field $\phi$ in the Einstein-frame.  Note that a plateau only occurs if $f(\phi)$ or $f(R)$ are precisely cutoff at quadratic order, when there is no reason why there should not be higher order terms.  Yet the addition of any one higher order term is enough to ruin plateau inflation.  

A third possibility is periodic potentials of the type shown in Fig.\,1c, as occurs for axion-like fields ({\it e.g.}, as in natural inflation \cite{Freese:1990rb} or in string theory moduli).  This form is enforced by symmetry to be periodic and, unlike the previous cases, forbidden to have power-law behavior at large $\phi$.  This makes it the best-case scenario for evading the unlikeliness problem.  The problem arises if there are any non-axion-like scalar fields that define the vacuum since they will generically have power-law behavior at large $\phi$.  The more ordinary scalar fields that exist in fundamental theory, the more avenues there are for power-law inflation, each of which is exponentially favored over plateau-like inflation from the periodic potential but disfavored by Planck2013. 

Hence, none of these three cases evades the unlikeliness problem.  At the same time, it is clear that none does anything to evade the new initial conditions problem caused by Planck2013. In each case, the plateau-like inflation begins well after the big bang, enabling kinetic and gradient energy to dominate right after the big bang.  

A fourth possibility consists of models, like those sketched in Fig.\,1d, in which complicated features are added for the purpose of turning an unlikely model into a likely one.  For example, we have already shown that the plateau side (solid red) in Fig.\,1a has exponentially less inflation than the power-law side and an initial conditions issue; so the fact that Planck2013 disfavors the power-law and favors the plateau is a problem.   By grafting the sharp upward bend or false vacuum (dotted green) onto the plateau in Fig.\,1d, the combination technically evades those problems, but at the expense of complicating the potential.  So, in terms of the addressing the central issue of this Letter -- does Planck2013 really favor the simplest inflationary model? -- this approach does not change the answer.

Furthermore, the only reason for grafting onto a plateau model rather than some other potential shape is because of the foreknowledge that the plateau model fits Planck2013 data.  That means, effectively, what was supposed to be predicted output of the model has now been used as an input in its design.  It does not make sense to apply the unlikeliness criterion to models in which the very same volume and initial conditions test criteria were already ``wired in'' as input.  In fact, not only has the likeliness criterion been used as input, but all the Planck2013 data (tilt, tensor modes, spatial curvature, non-Gaussianity) have been used in selecting to graft onto a plateau potential rather than some other shape potential. If the only way the inflationary paradigm will work is by delicately designing all the test criteria and data into the potential, this is trouble for the paradigm.

\section{More trouble for inflation from the LHC?}
Thus far, we have only focused on recent results from Planck2013, but recent measurements of the top quark and Higgs mass at the LHC and the absence of evidence for physics beyond the standard model could be a new source of trouble for the inflationary paradigm and big bang cosmology generally \cite{ATLAS:2012oga,Lai:2013yaa}.  Namely, the current data suggests that the current symmetry-breaking vacuum is metastable with a modest-sized energy barrier ($(10^{12}$ GeV)$^4$) protecting us from decay to a true vacuum with large negative vacuum density \cite{Degrassi:2012ry}.  This conclusion is speculative since it assumes no new physics for energies less than the Planck scale, which is unproven.  Nevertheless, this is the simplest interpretation of the current data and its consequences are dramatic; hence, we consider the implications here.

The predicted lifetime of the metastable vacuum is large compared to the time since the big bang, so there is no sharp conflict with observations.  The new problem is explaining how the universe managed to become trapped in this false vacuum whose barriers are tiny (by a factor of $10^{28}$!) compared to the Planck density when it is obviously much more probable for the field to lie outside the barriers than within them.  However, if the Higgs field lies outside the barrier, its negative potential energy density will tend to cancel the positive energy density of the inflaton and block inflation from occurring, unless one assumes large-field inflation and a certain kind of coupling between the inflaton and the Higgs \cite{Lebedev:2012sy,Kobakhidze:2013tn}.  Even in the unlikely case that the Higgs started off trapped in its false vacuum and inflation began, the inflaton would induce de Sitter-like fluctuations in all degrees of freedom that are light compared to the Hubble scale during inflation.  These tend to kick the Higgs field out of the false vacuum, unless the Hubble constant during inflation is smaller than the barrier height \cite{Espinosa:2007qp}.  Curiously, a way to evade the kick-out is if all inflation (not just the last 60 e-folds) occurs at low energies where the de Sitter fluctuations are smaller than the barrier height.  This  would be possible if the only possible inflaton potentials are plateau-like with sufficiently low plateaus: the very same potentials that have the initial conditions and multiverse problems.

\section{Discussion}
In testing the validity of any scientific paradigm, the key criterion is whether measurements agree with what is expected given the paradigm.  In the case of inflationary cosmology, this test can be divided into two questions: (A) {\it are the observations what is expected, given the inflaton potential X?}, here the analysis assumes classical slow-roll, no multiverse, and ideal initial conditions; and (B) {\it is the inflaton potential X that fits the data what is expected according to the internal logic of the paradigm?}.   In order to pass, {\it both} questions must be answered in the affirmative.  

The Planck2013 analysis, like many previous analyses of cosmic parameters, focused on Question A.  Based on tighter constraints on flatness, the power spectrum and spectral index, and non-Gaussianity, the conclusion from Planck2013 was that single-field plateau-like models are the simplest that pass and they pass with high marks.

However, our focus in this Letter has been Question B -- are plateau-like models expected, given the inflationary paradigm?  Based on the very same tightened constraints from Planck2013, we have identified three independent issues for plateau-like models: a dangerous new type of initial conditions problem, a twist on the multiverse problem, and, for the first time, an inflationary unlikeliness problem.  The fact that a single data set like Planck2013 can expose three new problems is a tribute to the quality of the experiment and serious trouble for the paradigm.
    
Future data can amplify, confirm, or diffuse the three problems.  Detecting tensor modes and constraining the non-Gaussianity to be closer to zero would ease the problems provided the $r$-$n_s$ values are consistent with a simple power-law potential.  Given the Planck2013 value for the tilt ($n_s = 0.9603 \pm 0.0073$), the only simple chaotic model that can be recovered is $m^2\,\phi^2$, predicting $0.13 \lesssim r \lesssim 0.16$ (depending on the value of $N$). 
Alternatively, if the observed $r$ lies at 0.01 or below, power-law models are ruled out and all three current problems remain.
Yet a third possibility is finding no tensor modes or detecting non-negligible non-Gaussianity ({\it e.g.,} $f_{\text{NL}} \sim 8$ is well within Planck2013 limits but inconsistent with plateau models); measurements like these would create yet more problems for the inflationary paradigm and encourage consideration of alternatives.

\section*{Acknowledgments}
We thank T. Baker, W. Jones, J. Kovac, J.-L. Lehners, D. Spergel, and B. Xue for helpful discussions.
This work was supported in part by the US Department of Energy grant DE-FG02-91ER40671 (PJS) and by NSF grant AST-0907890 and NASA grants NNX08AL43G and NNA09DB30A (AL). AI gratefully acknowledges the support of the Fritz Thyssen Foundation. PJS is grateful to the Simons Foundation and the Radcliffe Institute for providing support during his leave at Harvard and to the Institute for Theory and Computation at the Harvard-Smithsonian for hosting him during the period that this work was done.

\bibliographystyle{apsrev}
\bibliography{PlanckInflation}

\end{document}